\documentclass[pra,english,preprintnumbers,amsmath,amssymb,nofootinbib,twocolumn,superscriptaddress]{revtex4-1}

\usepackage[latin1]{inputenc}
\usepackage{graphicx}
\usepackage{bbm}
\usepackage{amssymb}
\usepackage{amsmath}
\usepackage{nicefrac}

\usepackage{dsfont}

\def\0#1#2{\frac{#1}{#2}}

\def\s0#1#2{\mbox{\small{$ \frac{#1}{#2} $}}}

\newcommand{\Tr}{\mathrm{Tr}}

\newcommand{\E}{\mathrm{e}}

\newcommand{\be}{\begin{eqnarray}}
\newcommand{\ee}{\end{eqnarray}}

\newcommand{\nn}{\nonumber }

\newcommand{\zgc}{\mathcal Z}

\usepackage{babel}
\makeatother
\begin{document}

\title{Phases of spin- and mass-imbalanced ultracold Fermi gases in harmonic traps}

\author{Jens Braun} 
\affiliation{Institut f\"ur Kernphysik (Theoriezentrum), Technische Universit\"at Darmstadt, 
D-64289 Darmstadt, Germany}
\affiliation{ExtreMe Matter Institute EMMI, GSI, Planckstra{\ss}e 1, D-64291 Darmstadt, Germany}
\author{Joaqu\'{\i}n E. Drut} 
\affiliation{Department of Physics and Astronomy, University of North Carolina, Chapel Hill, NC 27599, USA}
\author{Thomas Jahn}
\affiliation{Institut f\"ur Kernphysik (Theoriezentrum), Technische Universit\"at Darmstadt, 
D-64289 Darmstadt, Germany}
\author{Martin Pospiech}
\affiliation{Institut f\"ur Kernphysik (Theoriezentrum), Technische Universit\"at Darmstadt, 
D-64289 Darmstadt, Germany}
\author{Dietrich Roscher}
\affiliation{Institut f\"ur Kernphysik (Theoriezentrum), Technische Universit\"at Darmstadt, 
D-64289 Darmstadt, Germany}

\begin{abstract}
We analyze the phase structure of mass- and spin-imbalanced unitary Fermi gases in harmonic traps. To this end,
we employ Density Functional Theory in the local density approximation. Depending on the values of 
the control parameters measuring mass and spin imbalance, we observe that three regions exist
in the trap, namely: a superfluid region at the center, surrounded by a mixed region of resonantly
interacting spin-up and spin-down fermions, and finally a fully polarized phase surrounding the previous two regions. 
We also find regimes in the phase diagram where the existence of a superfluid region at the center of the trap
is not energetically favored. We point out the limitations of our approach at the present stage, and 
call for more detailed ({\it ab initio}) studies of the equation of state of uniform, mass-imbalanced unitary 
Fermi gases.
\end{abstract}

\maketitle

%
\section{Introduction}\label{sec:intro}
Ultracold Fermi gases have attracted a great deal of attention from a variety of research fields in the past 
15 years. This interest can be traced back to the fact that quantum many-body phenomena, such as 
Bardeen-Cooper-Schrieffer~(BCS) superfluidity and Bose-Einstein condensation~(BEC) can be studied experimentally 
with very high precision in some cases~\cite{PhysRevLett.106.215303,2012NatPh...8..366V,PhysRevLett.110.055305} 
(see Ref.~\cite{revexp} for a review) which opens up the possibility to test our theoretical understanding of such phenomena in 
a very clean way~\cite{Bloch:2008zzb,Giorgini:2008zz}. 

From an experimental point of view, the control parameters are the density~$n$ and the
s-wave scattering length~$a_{\rm s}$, provided that the (effective) range of the interaction can be neglected. The latter 
is true for a sufficiently dilute Fermi gas to a very good approximation. 
The dynamics of the system is then entirely controlled by the dimensionless
parameter~$n^{\frac{1}{3}}a_{\rm s}$. A particularly interesting limit is the so-called unitary regime, which is
characterized by~$a_{\rm s}\to \infty$. In the presence of a Feshbach resonance, 
the latter {can be tuned} with the aid of an external magnetic field which, however, is currently only possible 
for a limited number of (meta)stable atoms in the nuclear chart, such as ${}^6$Li and~${}^{40}$Ka.

In the present work, we shall restrict ourselves to the unitary regime defined above. In this limit, the 
only scale left in the problem is the density~$n$, at least for a uniform system. For trapped gases, as realized 
in experiments, an additional length scale enters the problem, namely the one associated with the (harmonic) trap
potential. This scale affects the dynamics of the system and may therefore alter
the phase structure compared to the uniform system. Studies of such finite-size effects are of utmost importance
to better connect our theoretical understanding of quantum many-body phenomena with experiment. 

Studies of unitary Fermi gases, even in the absence of an external potential, are already hampered by the fact that 
a small expansion parameter remains to be identified, which makes the use of non-perturbative tools 
unavoidable~\cite{ZwergerBook}. For the case of spin- and mass-imbalanced Fermi gases, which
are the focus of this work, even less is known beyond
the mean-field approximation, although great efforts have been made in recent years to study  
mass-imbalanced (see, e.g., Refs.~\cite{CRLC,Gezerlis:2009xp,Gandolfi:2010,2010PhRvA..82a3624B,BaarsmaStoof}) as well as
spin-imbalanced (see, e.g. Refs.~\cite{Chevy:2006,Lobo:2006,BulgacForbes,Chevy,KBS,Schmidt:2011zu}) unitary Fermi gases. 
We refer the reader to Refs.~\cite{ChevyMora,StoofGubbels} for more general reviews. We note, however, that {\it ab initio} 
studies of mass- and spin-imbalanced Fermi gases are generally out of reach for (lattice) Monte Carlo (MC) calculations due
to the appearance of a sign problem~\cite{Drut:2008}, which calls for the development and use of novel 
techniques~\cite{Braun:2012ww,Roscher:2013aqa}. Finally, the consideration of trap effects represents an additional 
(technical) complication for {\it ab initio} studies in general, especially at finite temperature.

In order to study trapped, three-dimensional unitary Fermi gases, Density Functional Theory (DFT) provides a viable 
framework (see, e.g., Refs.~\cite{DreizlerGross,EngelDreizler} for an introduction). In principle, DFT allows for an exact
solution of a given many-body problem. In practice, however, DFT studies rely on an approximation of the full
energy density functional. The simplest form is the so-called local density approximation (LDA), which represents
the zeroth order of an expansion of the energy density functional in terms of gradients of the density.
The density functional is then given by the volume integral over the uniform equation of state with the uniform 
densities replaced by their space-dependent counterparts.
For trapped unitary Fermi gases, LDA studies turn out to be quite successful, at least for systems with
many atoms. In fact, they show even quantitative agreement with experiments in some cases (see, e.g., 
Refs.~\cite{2006PhRvL..97u0402G,Bulgac:2007ah,RLS,2008PhRvA..78f3602H}). 
Ultimately, the predictive power of LDA studies depends strongly on the quality 
of the employed equation of state of the uniform system.

In the present work, we aim to understand the dynamics and phase structure of trapped
unitary, mass- and spin-imbalanced Fermi gases, including the computation of density profiles. 
To this end, we construct an energy density functional in LDA following Ref.~\cite{RLS}. 
Studies of the phase structure of mass-imbalanced unitary Fermi
gases represent a comparatively new field from the experimental perspective~\cite{PhysRevLett.100.053201,PhysRevLett.102.020405,%
PhysRevLett.106.115304,2011EPJD...65..223R,2012Natur.485..615K}.
Therefore, even though the accuracy of our predictions is limited, our present study may still provide useful insights 
into the dynamics of trapped spin- and mass-imbalanced Fermi gases.
For instance, it may help improve future DFT studies of such systems (going beyond the LDA), and it may provide 
guidance for experiments aiming at a study of the phase structure for non-vanishing spin- and mass-imbalance.

The present work is organized as follows: In Sect.~\ref{sec:formalism}, we give a detailed discussion of the
formalism underlying our studies, including a brief discussion of the uniform system. Our results for the phase
structure of a trapped spin- and mass-imbalanced unitary Fermi gas are then presented in Sect.~\ref{sec:res}. Our
summary is found in Sect.~\ref{sec:sum}.

\section{Formalism}\label{sec:formalism}
\subsection{Uniform System}\label{sec:formalismuf}
For a uniform system, the partition function of a unitary Fermi gas reads
\be
&&\zgc (T,m_{\uparrow},m_{\downarrow},{\mu},{h})\nn\\
&&\qquad\quad\qquad=\Tr\left[ \E^{-\beta(\hat{H}-{\mu}(\hat{N}^{}_{\uparrow}+\hat{N}^{}_{\downarrow})
-{h} (\hat{N}^{}_{\uparrow}-\hat{N}^{}_{\downarrow}))}\right]\,, \label{eq:Z} 
\ee
where $\beta=1/T$ is the inverse temperature. The Hamiltonian~$\hat{H}$ describes 
the dynamics of a theory with only two fermion species, denoted by \mbox{$\uparrow$ and $\downarrow$}, interacting only
via a zero-range two-body interaction:
\be
&&\hat{H}\!=\!
\int d^{3}x \Bigg [\!
 \sum_{\sigma=\uparrow,\downarrow}
  \hat{\psi}_{\sigma}^{\dagger}(\mathbf{x})\left(\frac{-\vec{\nabla}^2}{2m_{\sigma}}\right)\hat{\psi}_{\sigma} (\mathbf{x}) 
+ \bar{g} \hat{\rho}_\uparrow (\mathbf{x}) \hat{\rho}_\downarrow (\mathbf{x}) \Bigg]\,.\nn
\ee
The operators~$\hat{\rho}^{}_{\uparrow,\downarrow}$ are the particle density operators associated with the two 
fermion species, and $\hat{N}^{}_{\uparrow,\downarrow}$ are the corresponding particle-number operators. 
In order to study the unitary regime ($a_{\rm s}\to\infty$), the coupling~$\bar{g}$ must be chosen accordingly.

The masses of the two species are given by~$m_{\uparrow}$ and~$m_{\downarrow}$, respectively. 
Moreover, we have introduced the average chemical potential~${\mu}=(\mu^{}_{\uparrow}+\mu^{}_{\downarrow})/2$ 
and the asymmetry parameter~${h}=(\mu^{}_{\uparrow}-\mu^{}_{\downarrow})/2$. The corresponding dimensionless measure
for the spin imbalance of the system is given by
\be
\bar{h}=\frac{h}{\mu}=\frac{\mu_{\uparrow}-\mu_{\downarrow}}{\mu_{\uparrow}+\mu_{\downarrow}}\,.\label{eq:hbdef}
\ee
Along these lines it is also convenient to introduce a measure for the mass imbalance of the system:
\be
m_{+}=\frac{4 m_{\uparrow}m_{\downarrow}}{m_{\uparrow}+m_{\downarrow}}\,,\quad
m_{-}=\frac{4 m_{\uparrow}m_{\downarrow}}{m_{\downarrow}-m_{\uparrow}}\,,\quad
\bar{m}=\frac{m_{+}}{m_{-}}\,.
\ee
The parameter~$\bar{m}$ measures the relative strength of the mass imbalance where~$-1 < \bar{m} < 1$.
At this point, we would like to emphasize that the theory is invariant under the following simultaneous transformation 
of~$\bar{h}$ and~$\bar{m}$: $\bar{h}\to -\bar{h}$ and~$\bar{m}\to-\bar{m}$. To fix the scales, we henceforth 
set \mbox{$m_{+}=1$} corresponding to $2m=1$ for $m_{\uparrow}=m_{\downarrow}=m$.

\begin{figure}[t]
\includegraphics[width=0.95\columnwidth]{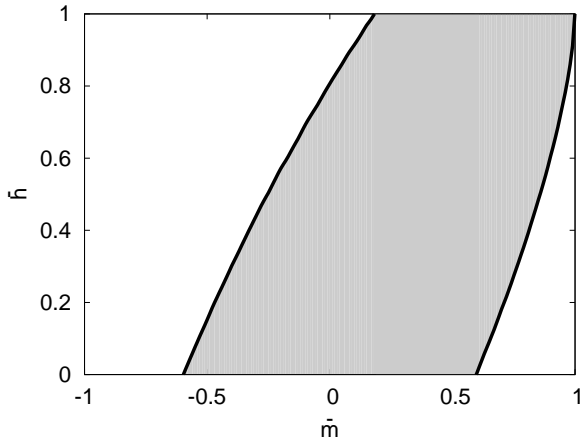}
\caption{\label{fig:1}Phase diagram of a mass- and spin-imbalance unitary Fermi gas as obtained from a 
mean-field approximation. The black solid lines mark the first-order phase transitions between the BCS-type superfluid
phase (gray-shaded area) associated with a spontaneous breakdown of the U($1$) symmetry of the theory and the 
non-superfluid phase associated with ungapped fermions and restored U($1$) symmetry.
}
\end{figure}

In the mean-field approximation, the 
zero-temperature phase diagram of the uniform system can now be computed straightforwardly, provided that we do not take
into account the possibility of the existence of inhomogeneous phases~\cite{BaarsmaStoof,Roscher:2013cma} (see Appendix~\ref{app:mf} 
for details). The result of such a mean-field study is shown in Fig.~\ref{fig:1}. We find that, for any value of the spin-imbalance
parameter $\bar{h}$, the system assumes a ground state associated with a BCS-type superfluid, provided the mass-imbalance
parameter $\bar{m}$ is chosen accordingly. The shape of the BCS-type phase can be understood in simple terms: Increasing~$\bar{h}$ for a fixed value of~$\bar{m}$
induces a difference in the Fermi momenta $k_{{\rm F},\uparrow}$ and~$k_{{\rm F},\downarrow}$ associated with the two fermion
species:
\be
\frac{1}{\sqrt{\mu}}\left(k_{{\rm F},\uparrow}-k_{{\rm F},\downarrow}\right) 
=\sqrt{\frac{1+\bar{h}}{1+\bar{m}}}-\sqrt{\frac{1-\bar{h}}{1-\bar{m}}}\,.\label{eq:kfdiff}
\ee
Assuming that the emergence of a BCS-type ground-state requires the Fermi momenta of the two species to be approximately 
equal, we conclude that the difference of the Fermi momenta induced by an increase of~$\bar{h}$ can be compensated by a corresponding
increase of~$\bar{m}$.\footnote{Note that we have $k_{{\rm F},\uparrow}=k_{{\rm F},\downarrow}$ for $\bar{h}=\bar{m}$.} On the other hand, we find 
that values for~$\bar{m}$ exist for which the system does not assume a superfluid ground state, independent of our choice
for the spin-imbalance parameter~$\bar{h}$. This observation is also in accordance with our simple considerations based on Eq.~\eqref{eq:kfdiff}.
Moreover, our mean-field analysis suggests that, for~$\bar{m}\to 1$, the phase characterized by a superfluid BCS-type ground state is only
energetically favored in the limit~$\bar{h}\to 1$.
We shall come back to this observation below.

Finally, we comment on the value of the so-called {\it Bertsch} parameter~$\xi_\text{S}$ in the superfluid phase which is a
measure for the ground-state energy of the system. In the case of a mass- and
spin-imbalanced Fermi gas, this parameter can be defined as follows:
\be
\varepsilon_{\text{S}}\equiv \frac{E_{\text{S}}}{N_{\text{S}}}:=\frac{1}{2}\left(\varepsilon^{\text{S}}_{\uparrow}
+\varepsilon^{\text{S}}_{\downarrow}\right)\,, \label{eq:eossf0}
\ee
where
\be
\varepsilon^{\text{S}}_{\uparrow,\downarrow}  \equiv \frac{E^{\text{S}}_{\uparrow,\downarrow}}{N^{\text{S}}_{\uparrow,\downarrow}} 
&:=& \frac{3}{5} \xi_\text{S} \frac{1}{2 m_{\uparrow,\downarrow}} \left( 6 \pi^2 n_\text{S} \right)^{\frac{2}{3}} \nn\\
&\phantom{:}=& \frac{3}{5} \xi_\text{S} \left( 1 \pm \bar{m} \right) \left( 6 \pi^2 n_\text{S} \right)^{\frac{2}{3}}\,.
\label{eq:eossf}
\ee
Note that our conventions are such that~$n_{\text{S}}=n_{\uparrow}=n_{\downarrow}$ in the 
superfluid phase.\footnote{Thus, $n_{\rm s}$ 
should here not be confused with the so-called pair density in the superfluid phase, see also Ref.~\cite{RLS} 
where similar conventions have been used.
Moreover, we would like to add that, in our mean-field approximation, we indeed find 
that~$n_{\uparrow}=n_{\downarrow}$ in the BCS phase.} The quantities~$N_{\uparrow,\downarrow}$
denote the number of spin-up and spin-down fermions, respectively.
For~$\bar{m}=0$, this definition reduces to the standard definition of the 
{\it Bertsch} parameter. With our mean-field approach, we find that~$\xi_{\text{S}}$ is independent of~$\bar{m}$ and~$\bar{h}$.
In our DFT study presented below, we shall employ this observation to model the equation of state which governs the dynamics
of the superfluid region in the trap.
\subsection{Trapped System}\label{sec:formalismtr}
In order to account for trap effects, we could in principle include terms in the underlying Hamilton operator which couple
(harmonic) external potentials~$V_{\uparrow,\downarrow}$ to the density operators~$\hat{\rho}_{\uparrow,\downarrow}$ and compute 
the partition function~$\mathcal Z$. Although such an approach represents a rigorous way for a study of the phase
diagram, the computation of this partition function~$\mathcal Z$ with {\it ab initio} approaches, such
as lattice Monte-Carlo simulations, appears to be currently out of reach. Here, we therefore use DFT in LDA following Refs.~\cite{RLS}. This allows us to include and study trap effects
in a very efficient way. However, it requires the knowledge of the equation of state of the uniform system. 

As the precise determination of the equation of state of a uniform spin- and mass-imbalanced unitary Fermi gas is itself a highly 
challenging and unsolved problem, we shall utilize results from various methods to model it in the following.

For the energy density functional~$E$ underlying our study, we shall use the following ansatz:
\be
&& E[n_\text{S},n_\uparrow,n_\downarrow] = 2 \int\limits_{r \leq R_\text{S}}^{} \mathrm{d}^3\mathrm{r} \ n_\text{S} \left[ \varepsilon_{\text{S}} - \mu_\text{S}^0 
+ \frac{1}{2} \left( V_\uparrow + V_\downarrow \right) \right]\nn\\
&& \qquad\qquad\qquad\quad + \int\limits_{R_\text{S} \leq r \leq R_{\text{max.}}}^{} \!\!\!\!\!\mathrm{d}^3\mathrm{r} 
\left[ \varepsilon_\text{N}(x) n_\uparrow + V_\uparrow n_\uparrow + V_\downarrow n_\downarrow\right.\nn\\
&& \qquad\qquad\qquad\qquad\qquad\qquad\qquad  
\left. - \mu_\uparrow^0 n_\uparrow - \mu_\downarrow^0 n_\downarrow \right]\,,\label{eq:edf}
\ee
where $\varepsilon_{\text{S}}\equiv \varepsilon_{\text{S}}(n_{\text{S}}(\vec{r}\,))$ and~$\varepsilon_\text{N}(x)$ 
with $x\equiv x(\vec{r})=n_{\downarrow}(\vec{r}\,)/n_{\uparrow}(\vec{r}\,)$
are the equations of state of the superfluid and the normal phase, respectively.
The (isotropic) trap potentials are given by
\be
V_{\uparrow,\downarrow}\equiv V_{\uparrow,\downarrow}(\vec{r}\,) =
\frac{1}{2} m_{\uparrow,\downarrow} \omega_{\uparrow,\downarrow}^2 \vec{r}^{\,2} 
\equiv \frac{1}{4} \frac{1}{1 \pm \bar{m}} \omega_{\uparrow,\downarrow}^2 \vec{r}^{\,2}\,.
\ee
For convenience, we set~$\omega_{\downarrow}=\alpha\omega_{\uparrow}$ with~$\alpha$ being a measure
for the difference in the trap potentials for the spin-up and spin-down fermions, respectively. 
The quantity~$R_{\text{S}}$ in Eq.~\eqref{eq:edf} determines the radial extent of the superfluid core in the center of the trap, whereas the quantity~$R_{\text{max.}}$
determines the radius of the total system.\footnote{Without loss of generality, we tacitly assume that $N_{\uparrow}\geq N_{\downarrow}$ in the following.} 
In our ansatz for the functional~$E$, we
have also included chemical potentials~$\mu_{\uparrow}^{0}$ and~$\mu_{\downarrow}^{0}$ for the spin-up and spin-down species, respectively. 
They fix the chemical potential~$\mu_{\text{S}}^{0}=(\mu_{\uparrow}^{0}+\mu_{\downarrow}^{0})/2$ associated with the superfluid region as
we assume chemical equilibrium between the superfluid core~($0\leq R\leq R_{\text{S}}$) and the surrounding 
normal region~($R_\text{S} < r \leq R_{\text{max.}}$). 

In order to use the energy density functional~\eqref{eq:edf} to compute the phase structure of trapped spin- and mass-imbalanced
Fermi gases, we finally need to specify the equation of state of the superfluid phase and the normal phase. In LDA, the latter are given by the corresponding
equations of state of the uniform system by replacing the uniform densities with space-dependent densities, i.e. $n_{\text{S}}\to n_{\text{S}}(\vec{r})$ 
and~$n_{\uparrow,\downarrow}\to n_{\uparrow,\downarrow}(\vec{r})$ (see, e.g., Refs.~\cite{DreizlerGross,EngelDreizler}). For the superfluid region,
the equation of state can then be conveniently parameterized by the {\it Bertsch} parameter~$\xi_{\text{S}}$ (see Eq.~\eqref{eq:eossf0}).
For a mass- and spin-balanced unitary Fermi gas, the {\it Bertsch} parameter has been computed 
many times with the aid of MC techniques  
{and found to be~$\xi_{\text{S}}\approx 0.42$~\cite{PhysRevLett.91.050401,PhysRevLett.93.200404}. 
More recent estimates place $\xi_{\text{S}}$ at about 0.375 (see e.g.~\cite{Endres:2012cw} and references therein) 
but for reasons that are clarified below we use the previous value,\footnote{Recall that our conventions are such 
that~$n_{\text{S}}=n_{\uparrow}=n_{\downarrow}$ which result in a factor of two in 
front of the integral associated with the superfluid equation of state in Eq.~\eqref{eq:edf}.}
even in the case of finite 
spin- and mass-imbalance. Recall that our mean-field study in Sect.~\ref{sec:formalismuf} 
indeed suggests that this parameter does not depend on~$\bar{h}$ and~$\bar{m}$
in the superfluid phase.}
\begin{figure*}[t]
\includegraphics[width=0.95\columnwidth]{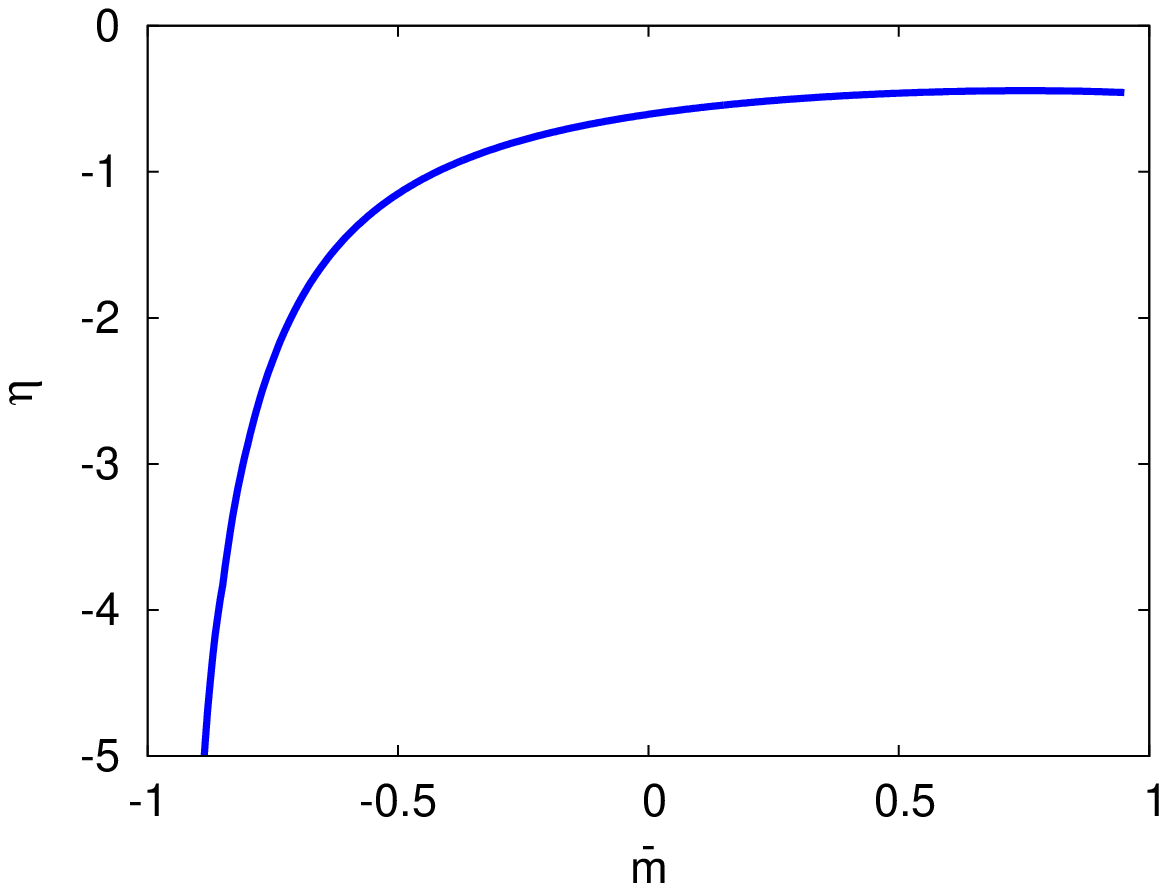}
\includegraphics[width=0.95\columnwidth]{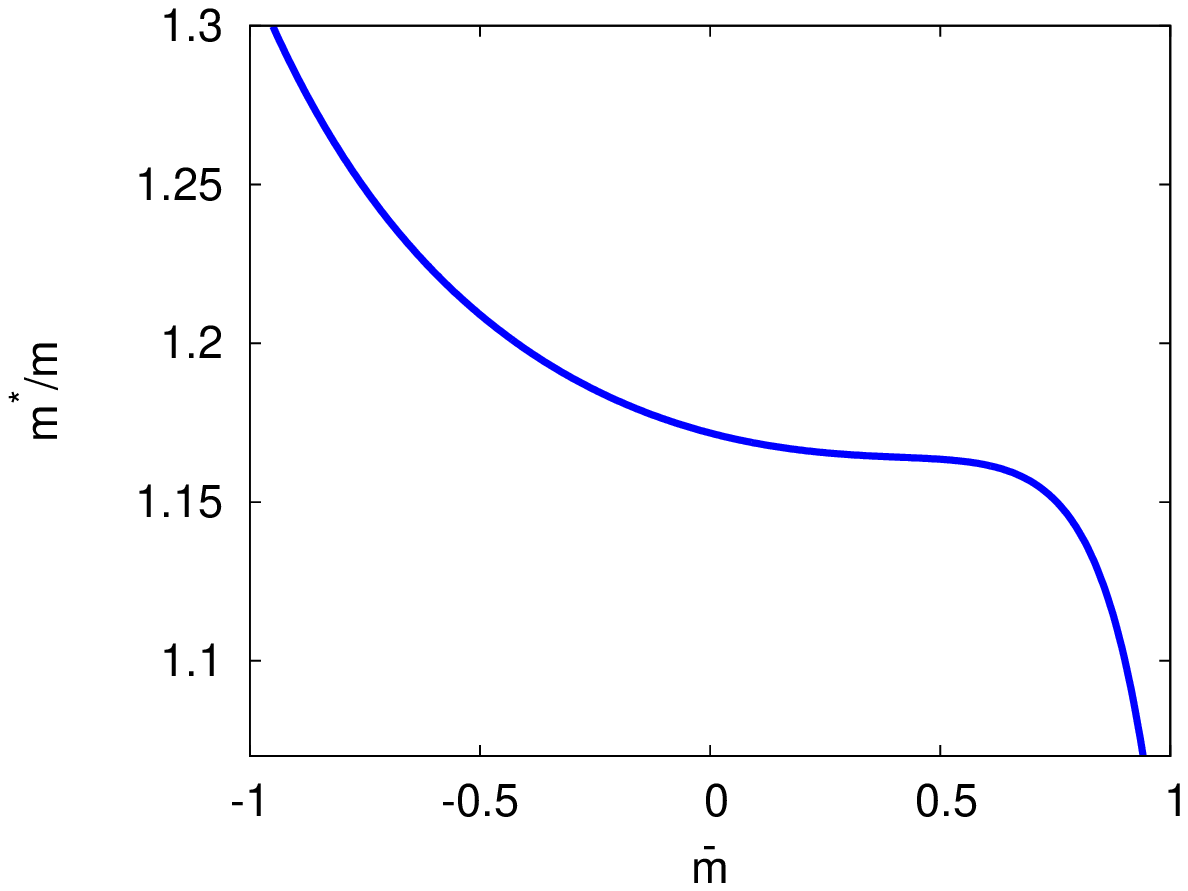}
\caption{\label{fig:2} Dependence of the energy gain~$\eta$ and the effective mass~$m^{\ast}$ on the relative mass difference~$\bar{m}$. For~$\bar{m}=0$, we find~$\eta\approx -0.60$
{and~$m^{\ast}/m\approx 1.17$. In} order to compute~$\eta$ and~$m^{\ast}$, we have used
a variational ansatz as described in Refs.~\cite{Chevy,Chevy:2006,CRLC}. Upon numerical errors, our results agree with those reported in Ref.~\cite{CRLC}.
}
\end{figure*}

For the equation of state of the normal phase, we shall employ an ansatz which essentially mimics an 
expansion of the system about $x=n_{\downarrow}/n_{\uparrow}=0$: 
\be
\varepsilon_\text{N}(x) \equiv \frac{E(x)}{N_\uparrow}\equiv \frac{3}{5}\varepsilon_{\text{F} \uparrow}\varepsilon(x)\,.\label{eq:eosn}
\ee
Here,
\be
\varepsilon_{\text{F} \uparrow}:=\left( 1 + \bar{m} \right) \left( 6 \pi^2 n_\uparrow \right)^{\frac{2}{3}}
\ee
is the equation of state of a non-interacting system of spin-up fermions and
\be
\varepsilon(x):=1 +\frac{5}{3}\eta x + \frac{m}{m^\ast} x^{\frac{5}{3}} + Bx^2\label{eq:epsans}
\ee
determines the deviation from the equation of state of non-interacting spin-up fermions in the presence of spin-down fermions~\cite{Lobo:2006}.
Clearly, our model ansatz for $\varepsilon (x)$ should be considered as an approximation of the full equation of state. The parameters~$\eta$, $m^{\ast}$,
and~$B$ can be computed for the uniform system. The parameter~$\eta$ describes the gain of energy when a spin-down fermion is added to a sea of spin-up fermions.
The third term~$(\sim x^{5/3})$ describes the quantum pressure of a Fermi gas of quasi-particles with an effective mass~$m^{\ast}$. The parameter~$B$ is a measure
of the interaction between the quasi particles.
For~$\bar{m}=0$, these parameters have been computed with various different approaches~\cite{Chevy,Lobo:2006,Chevy:2006,CRLC,Bulgac:2007,PS:2007,PG:2007,Schmidt:2011zu}
and it has been found that the parameterization~$\eqref{eq:epsans}$ models very well results from MC simulations, even for large values of~$x$ (see Ref.~\cite{Lobo:2006}). {In our analysis we shall use the parameter values 
determined in those references, which rely on a value of the Bertsch parameter that differs from the latest estimate $\simeq 0.375$ by about 
10-20\%; however, we do not expect our results to be accurate at that level of 
precision. Clearly an updated analysis of the equation of state is called for, but is beyond the scope of this work.}

For~$\bar{m}\neq 0$, the parameters depend on~$\bar{m}$ and terms of higher order in $x$ may now become relevant, at least for large values of~$x$.
We shall come back to this point below. In any case, compared to the mass-balanced case, little is known about the precise dependence of the presently included parameters on~$\bar{m}$. Indeed, lattice MC simulations suffer from the so-called
sign-problem in this regime and therefore cannot be applied straightforwardly, which makes the use of new techniques indispensable in future studies~\cite{Roscher:2013aqa}. 
In the present work, we shall employ the results for the $\bar{m}$-dependence of~$\eta$ and~$m^{\ast}$ as obtained from a variational approach~\cite{CRLC}, which
have been found to be in good agreement with the commonly accepted values for~$\bar{m}=0$ (see Fig.~\ref{fig:2}).  
{For the parameter~$B$, we choose~$B=0.14$ in our present study, independent of~$\bar{m}$. For~$\bar{m}=0$, this 
corresponds to the value used in Ref.~\cite{RLS}.}
However, we have checked the robustness of our results for the 
phase structure by varying this parameter (see also our discussion in the next section). 

Having specified the equations of state of the superfluid and normal phases, the ground state of the trapped system is then 
obtained by minimizing the energy density functional~\eqref{eq:edf}
with respect to the densities~$n_{\text{S}}$, $n_{\uparrow}$, and~$n_{\downarrow}$, as well as with respect to the radius~$R_{\text{S}}$ of the superfluid phase:
\be
\frac{\delta E}{\delta n_\text{S}} = \frac{\delta E}{\delta n_\uparrow} = \frac{\delta E}{\delta n_\downarrow} = \frac{\partial E}{\partial R_{\text{S}}}\stackrel{!}{=}0\,. 
\ee
Note that the variation with respect to~$R_{\text{S}}$ ensures mechanical equilibrium between the superfluid and the surrounding normal phase. 

From the variation of~$E$ 
with respect to the densities, we obtain the following set of equations:
\be
\mu_\text{S}^0 &=& \xi_\text{S} \left( 6 \pi^2 n_\text{S} \right)^{\frac{2}{3}} + \frac{1}{2} \left( V_\uparrow + V_\downarrow \right)\,,\label{eq:mu0}\\
\mu_\uparrow^0 &=& \left( 1+\bar{m} \right) \left( 6 \pi^2 n_\uparrow \right)^{\frac{2}{3}} \delta(x) + V_\uparrow\,,\label{eq:muup}\\
\mu_\downarrow^0 &=& \frac{3}{5} \varepsilon'(x)\left( 1+\bar{m} \right) \left( 6 \pi^2 n_\uparrow \right)^{\frac{2}{3}} + V_\downarrow\,,\label{eq:mudo}
\ee
where~$\varepsilon'(x)\equiv\partial \varepsilon(x)/\partial x$ and
\be
\delta(x)=\varepsilon(x) - \frac{3}{5} x \varepsilon'(x)\,.
\ee
From Eqs.~\eqref{eq:mu0}-\eqref{eq:mudo} together with the constraint $\partial E/\partial R_{\text{S}}=0$, we find
\be
&&\varepsilon(x(R_\text{S})) + \frac{3}{5} \varepsilon'(x(R_\text{S})) \left( 1 - x(R_\text{S}) \right)\nn\\
&& \qquad\qquad\qquad - \left( \frac{2 \xi_\text{S}}{1+\bar{m}} \right)^{\frac{3}{5}} \varepsilon(x(R_\text{S}))^{\frac{2}{5}} = 0\,, \label{eq:xRs}
\ee
which determines the ratio~$x$ of the spin-down and spin-up density 
at the boundary between the superfluid and the partially polarized normal region. Note that this equation
does not depend on the trap parameter~$\alpha=\omega_{\downarrow}/\omega_{\uparrow}$. 

The radial extent of the cloud of the majority and minority fermions is implicitly defined 
by~$n_{\uparrow}(R_{\uparrow})=0$ and~$n_{\downarrow}(R_{\downarrow})=0$, respectively.
From Eq.~\eqref{eq:muup} and Eq.~\eqref{eq:mudo}, we obtain:
\be
R_\uparrow = \frac{2}{\omega_{\uparrow}}\sqrt{\mu_{\uparrow}^0\left( 1 + \bar{m} \right)}
\ee
and
\be
R_\downarrow =   \frac{2}{\omega_{\uparrow}}\sqrt{ \frac{\mu_\downarrow^0 - \frac{3}{5} \frac{\varepsilon'(0)}{\delta(0)} \mu_\uparrow^0}{\frac{1}{1-\bar{m}} \alpha^2 - \frac{3}{5} \frac{\varepsilon'(0)}{\delta(0)} \frac{1}{1 + \bar{m}}}}\,.
\ee
Keeping~$\omega_{\uparrow}$ fixed, we find that $R_{\downarrow}$ decreases for increasing~$\alpha$ as expected. The radius~$R_{\text{max.}}$ of the
total system is then given by~$R_{\text{max.}}=\max\{R_{\uparrow},R_{\downarrow}\}$.

In the present work, we aim at understanding the dynamics of the trapped system under a variation of the 
mass-imbalance parameter~$\bar{m}$ and the polarization~$P$,
\be
P=\frac{N_\uparrow - N_\downarrow}{N_\uparrow + N_\downarrow}\,,
\ee
which corresponds to the parameter~$\bar{h}$ in the uniform case.
In particular, we are interested in the computation of the so-called critical polarization~$P_{\rm c}(\bar{m})$,
above which the superfluid core ceases to exist. To this end, we first solve Eq.~\eqref{eq:xRs} for~$x(R_{\text{S}})$
for a given value of~$\bar{m}$. For a given value of~$R_{\text{S}}$ and~$n_\uparrow (R_\text{S})$, we can then compute the
chemical potentials~$\mu_{\uparrow}^{0}$ and~$\mu_{\downarrow}^{0}$. The knowledge of the latter enables us to determine the density 
profiles~$n_\uparrow(r)$ und $n_\downarrow(r)$ in the normal phase, i.e. for $r=|\vec{r\,}|\geq R_{\text{S}}$. Finally, the density
profile~$n_{\text{S}}(r)$ associated with the {superfluid region can be computed} with the aid of Eq.~\eqref{eq:mu0}. Note that, in our
present LDA study, the particle numbers $N_\uparrow$ and $N_\downarrow$ depend on our choice for~$n_{\uparrow}(R_{\text{S}})$ but
the critical polarization~$P_{\rm c}$ does not.
\section{Results}\label{sec:res}
\subsection{Mass-balanced, spin-imbalanced case}
In this section we briefly review the phase structure of the trapped 
spin-imbalanced unitary Fermi gas with~$\bar{m}=0$. Solving Eq.~\eqref{eq:xRs} in this case, we find
\be
x(R_\text{S}) \approx 0.54\,.
\ee
With $x(R_\text{S})$ at hand, we can now compute the density profiles and eventually the critical 
polarization~$P_\text{c}$:\footnote{Setting~$B=0$, we find~$x(R_{\text{S}})\approx 0.93$ and~$P_{\text{c}}\approx 0.30$. The strong
dependence of~$x(R_{\text{S}})$ and~$P_{\text{c}}$ on the value of~$B$ for~$\bar{m}=0$ indicates the relevance of the parameter~$B$
measuring the interaction between the quasi particles.}
\be
P_\text{c}(\bar{m}=0) \approx 0.68\,.
\ee
For~$P>P_\text{c}$, we find that there is no superfluid phase anymore in the center of the trap and we are left with
two distinct regions, namely a non-superfluid mixed region with~$n_{\uparrow}(r)\neq 0$ and~$n_{\downarrow}(r)\neq 0$~($R_\text{S}<r<R_{\downarrow}$)
and a fully polarized normal region with~$n_{\uparrow}(r)\neq 0$ and~$n_{\downarrow}(r)\equiv 0$~($R_\downarrow<r<R_{\uparrow}$). For~$P<P_\text{c}$, on the 
other hand, we have in addition a superfluid region in the center of the trap with~$n_\text{S}\neq 0$~($r<R_\text{S}$) (see also our discussion of density profiles below).

Our results for~$\bar{m}=0$ should be compared with the results of Ref.~\cite{RLS}. In the latter work, the authors have
found that~$x(R_\text{S}) \approx 0.44$ and~$P_\text{c} \approx 0.77$. These results as well as those for the density
profiles have been found to be in very good agreement with data from MIT experiments~\cite{2006PhRvL..97c0401S,2008Natur.451..689S}. 
For example,~$P_{\text{c}}\approx 0.75$ has been extracted from the experimental data.
{The discrepancy with our results can be traced back to the fact that we use $m^{\ast}/m=1.17$ for~$\bar{m}=0$ in 
the ansatz~\eqref{eq:epsans} instead of~$m^{\ast}/m=1.09$ as obtained from a MC study of the mass-balanced case~\cite{PG:2007}. 
In any case, we have checked that our results agree with those from Ref.~\cite{RLS} in the limit~$\bar{m}=0$, provided
we also use~$m^{\ast}/m=1.09$. For our studies of the full phase diagram in the $(P,\bar{m})$ plane, this implies that we 
underestimate the critical polarization~$P_\text{c}(\bar{m})$ by about~$10\%$, at least in the limit of small mass 
imbalance,~$|\bar{m}|\ll 1$.}
\begin{figure}[t]
\includegraphics[width=0.95\columnwidth]{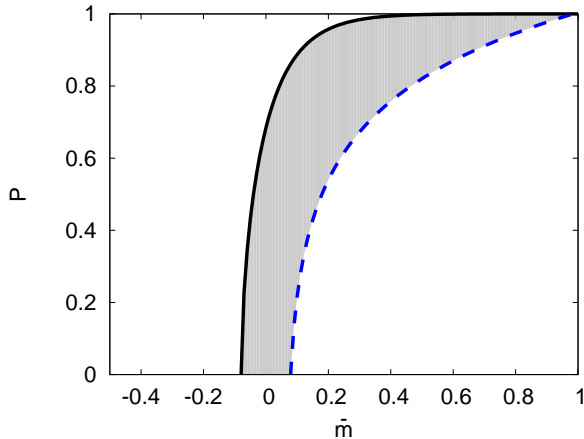}
\caption{\label{fig:3}(color online) 
Phase diagram of a spin- and mass- imbalanced unitary Fermi gas in the plane spanned by the polarization $P$ and the mass imbalance~$\bar{m}$ for~$\alpha=1$. The black solid line depicts the critical polarization~$P_{\text{c},<}(\bar{m})$ below which a superfluid region in the center of the trap is energetically favored. We find that a finite domain in this phase diagram exists (gray-shaded area) in which the center of the trap is governed by a superfluid region. Whereas the black solid line is a result of our DFT study, the existence of the (blue) dashed line follows from general considerations (see main text for a detailed discussion).
}
\end{figure}
%
\subsection{Mass- and spin-balanced case}
\subsubsection{Critical polarization as a function of mass imbalance and trap asymmetry}

Let us now turn to the discussion of the phase diagram of trapped mass- and spin-imbalanced unitary Fermi gases 
(see Fig.~\ref{fig:3}).
Lowering the polarization~$P$ for a given~$\bar{m}$ starting from a fully polarized system ($P=1$), we find that the superfluid core disappears
at a critical polarization~$P_{\text{c},<}$. Moreover, we find that~$P_{\text{c},<}$ increases with increasing~$\bar{m}$, 
\be
\frac{dP_{\text{c},<}(\bar{m})}{d\bar{m}}>0\,,
\ee
and that~$P_{\text{c},<}$ tends to zero for~$(\bar{m}-\bar{m}_{\text{c},<})\to 0^{+}$ 
where~$\bar{m}_{\text{c},<}\approx -0.08$.\footnote{Setting~$B=0$, we find~$\bar{m}_{\text{c},<} \approx -0.001$. This indicates again the relevance
of higher-order terms in~$x$ in our ansatz for the equation of state of the normal phase~\eqref{eq:epsans}.} 
For $\bar{m}<\bar{m}_{\text{c},<}$, 
there is no superfluid region in the center of the trap, independent of our choice for the polarization~$P$. For~$\bar{m}>\bar{m}_{\text{c},<}$ and~$P<P_{\text{c},<}(\bar{m})$,
we find that the existence of three distinct regions is energetically favored, namely a superfluid region in the center of the trap surrounded by a mixed region which is surrounded by
a fully polarized normal region (see also our discussion above for the case~$\bar{m}=0$).

\begin{figure}[t]
\includegraphics[width=0.95\columnwidth]{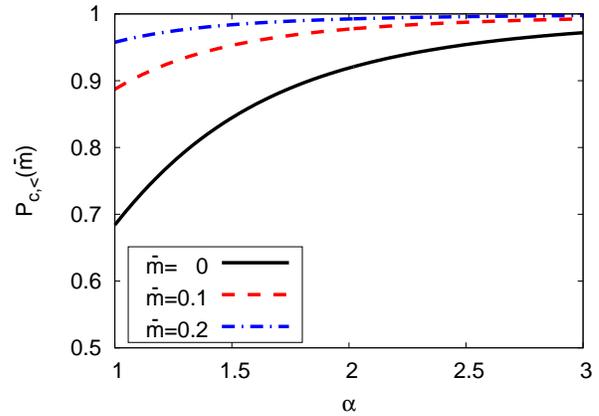}
\caption{\label{fig:3b}(color online) Dependence of the critical polarization~$P_{c,<}(\bar{m})$ on the trap-asymmetry parameter~$\alpha$ for $\bar{m}=0,0.1,0.2$.
}
\end{figure}

At this point, it is also interesting to discuss the effect of the trap-asymmetry parameter~$\alpha$. Increasing~$\alpha$ starting from~$\alpha=1$, we find that~$P_{\text{c},<}$ increases
as well (see Fig.~\ref{fig:3b}). However, we obtain that~$\bar{m}_{\text{c},<}\approx -0.08$ does not depend on~$\alpha$. 
Whereas the latter observation is an artifact of 
our ansatz for the energy density functional, the general observation that~$P_{\text{c},<}$ increases with~$\alpha$ for~$\bar{m}>0$ appears to be  
reliable. Indeed, from a physical point of view, this 
dependence of~$P_{\text{c},<}$ on~$\alpha$ can be traced back to the fact that the trap potential of the spin-down fermions becomes 
steeper when we increase~$\alpha$. Therefore the spin-down fermions are highly localized around the center of the trap and
their potential energy increases. For the spin-down fermions it is then energetically more favorable to form Cooper pairs with spin-up fermions and condense. This explains
the increase of the critical polarization with~$\alpha$.
\subsubsection{Superfluid region in the phase diagram}
\begin{figure}[t]
\includegraphics[width=0.95\columnwidth]{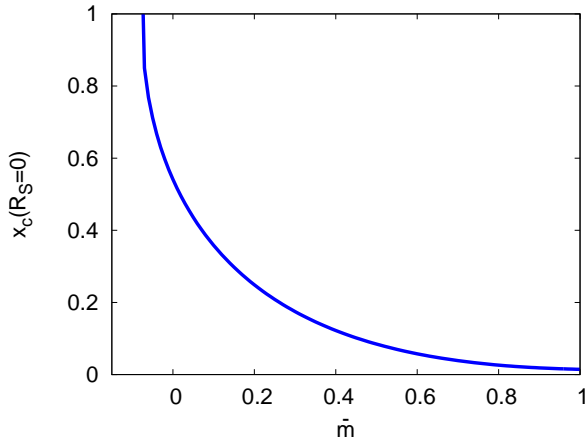}
\caption{\label{fig:4}(color online) Ratio of the spin-down and spin-up densities for~$r=R_{\text{S}}=0$ (i.e. for~$P=P_{\text{c},<}$) 
as a function of~$\bar{m}$. Recall
that~$R_{\text{S}}$ denotes the radius of the superfluid region which is zero for~$P=P_{\text{c},<}(\bar{m})$. 
Moreover, in our DFT study, $P_{\text{c},<}(\bar{m})$ is {defined to be the smallest}
value of the polarization for which $R_{\text{S}}$ becomes zero.}
\end{figure}

Next, we discuss the size of the region in the phase diagram in which a superfluid region at the core of the trap 
is energetically favored.
Applying our DFT approach to the regime with~$\bar{m}>\bar{m}_{\text{c},<}$ 
and~$P\ll P_{\text{c},<}(\bar{m})$, we find that a superfluid core is predicted to exist in the center of the trap for all values of~$0\leq P<P_{\text{c},<}(\bar{m})$ and~$\bar{m}_{\text{c},<}<\bar{m}<1$.
We now analyze this prediction with the aid of more general arguments.
To this end, we first recall that the uniform system is invariant under the simultaneous transformations $\bar{h}\to -\bar{h}$ and~$\bar{m}\to-\bar{m}$. For the trapped system, this translates
into an invariance under the simultaneous transformations~$P\to -P$ and~$\bar{m}\to -\bar{m}$, provided we consider the case~$\alpha=1$ for
the trap-asymmetry parameter.\footnote{Note that,
strictly speaking, there is no simple one-to-one relation between the parameter~$h$ and the polarization~$P$. 
However, both parameters are in principle related via a Legendre
transformation.} For~$\alpha\neq 1$, the system is invariant under the simultaneous transformations~$P\to -P$, $\bar{m}\to -\bar{m}$, 
and~$\omega_{\uparrow,\downarrow}\to \omega_{\downarrow,\uparrow}$.
In the {following we restrict ourselves} to the case~$\alpha=1$. With our symmetry consideration at hand, we then
expect that, in addition to the ``critical point"~$\bar{m}_{\text{c},<}$ at~$P=0$, a second ``critical point"~$\bar{m}_{\text{c},>}$ at~$P=0$ 
exists in the phase diagram with
\be
\bar{m}_{\text{c},>}=-\bar{m}_{\text{c},<}\,. \label{eq:mcp}
\ee
This implies that the regime in the phase diagram characterized by the existence of a superfluid core in the trap does not
extend to~$\bar{m}\to 1$ for~$P=0$, provided that our presently employed energy density functional has still predictive power in regions of the phase diagram
where~$x(r)=n_{\downarrow}(r)/n_{\uparrow}(r)\approx 1$. The latter is the case for~$\bar{m}<0$ and~$P\gtrsim P_{\text{c},<}(\bar{m})$ (see also Fig.~\ref{fig:4}).
However, only if our functional still provides an accurate description of the system for~$x(r)\lesssim 1$, it can be used to predict the existence of the point~$\bar{m}_{\text{c},<}$. 
We shall discuss the validity of our approach for~$\bar{m}<0$ in detail below. For the moment, we assume the existence of 
the second ``critical point"~$\bar{m}_{\text{c},>}$ which
is given by the intersection of the (blue) dashed line with the~$P=0$ axis in Fig.~\ref{fig:3}. In the following we shall 
refer to the associated line of critical polarizations as~$P_{\text{c},>}(\bar{m})$.
\subsubsection{Phase structure at large polarization and mass asymmetry}

To discuss the phase structure for large values of~$\bar{m}$ and~$P$, it is useful to consider again the phase diagram of the uniform system. In that case, we found that the size of the BCS-type phase shrinks to a single point for $\bar{m}\to 1$ (see 
Fig.~\ref{fig:1}). One may therefore be tempted to conclude that $\bar{m}\to 1$ necessarily implies
$P_{\text{c},>}(\bar{m})\to 1$.
For~$P<P_{\text{c},>}(\bar{m})$, we would then expect that only two distinct regions exist in the trap: a mixed phase surrounded by a 
fully polarized normal phase.
In Fig.~\ref{fig:3}, the (blue) dashed line depicts one possible functional form for~$P_{\text{c},>}(\bar{m})$ compatible with our general
considerations.

\begin{figure}[t]
\includegraphics[width=0.95\columnwidth]{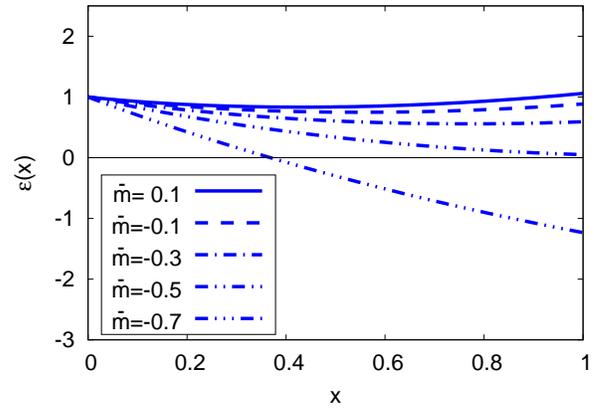}
\caption{\label{fig:5}(color online) 
Normalized equation of state~$\varepsilon(x)=(5/3)(\varepsilon_\text{N}(x)/\varepsilon_{\text{F} \uparrow})$
of the normal phase for various values of the mass-imbalance parameter~$\bar{m}$ as a function of~$n_{\downarrow}/n_{\uparrow}$, 
see also Eq.~\eqref{eq:epsans}. For~$\bar{m}\lesssim -0.5$, we observe 
that~$\varepsilon(x)$ becomes negative for a finite range of values of~$x$ indicating the breakdown of our ansatz for~$\varepsilon(x)$ 
for these values of~$\bar{m}$.
} 
\end{figure}

A word of caution should be added at this point: The parameters~$\bar{h}$ and~$P$ are related via 
a Legendre transformation which implies that there is indeed no straightforward mapping between the phase boundaries of the
BCS-type phase of the uniform system onto the lines of critical polarizations~$P_{\text{c},>}(\bar{m})$ and~$P_{\text{c},<}(\bar{m})$ of
the trapped system. Note that the constraint~$P_{\text{c},>}(\bar{m}_{\text{c},>})=0$ is not affected by this. However,
it may be the case that~$P_{\text{c},>}(\bar{m})<1$ for~$\bar{m}\to 1$. Therefore 
the functional form of~$P_{\text{c},>}(\bar{m})$ could very well be different from the one depicted in Fig.~\ref{fig:3}.
In any case, it follows that our more general discussion of the phase structure is in {contradiction with the results}
from our DFT study for~$\bar{m}>\bar{m}_{\text{c},<}$ and~$P\ll P_{\text{c},<}(\bar{m})$.
Basically, this observation allows for three different conclusions:
\begin{figure}[h!]
\includegraphics[width=0.95\columnwidth]{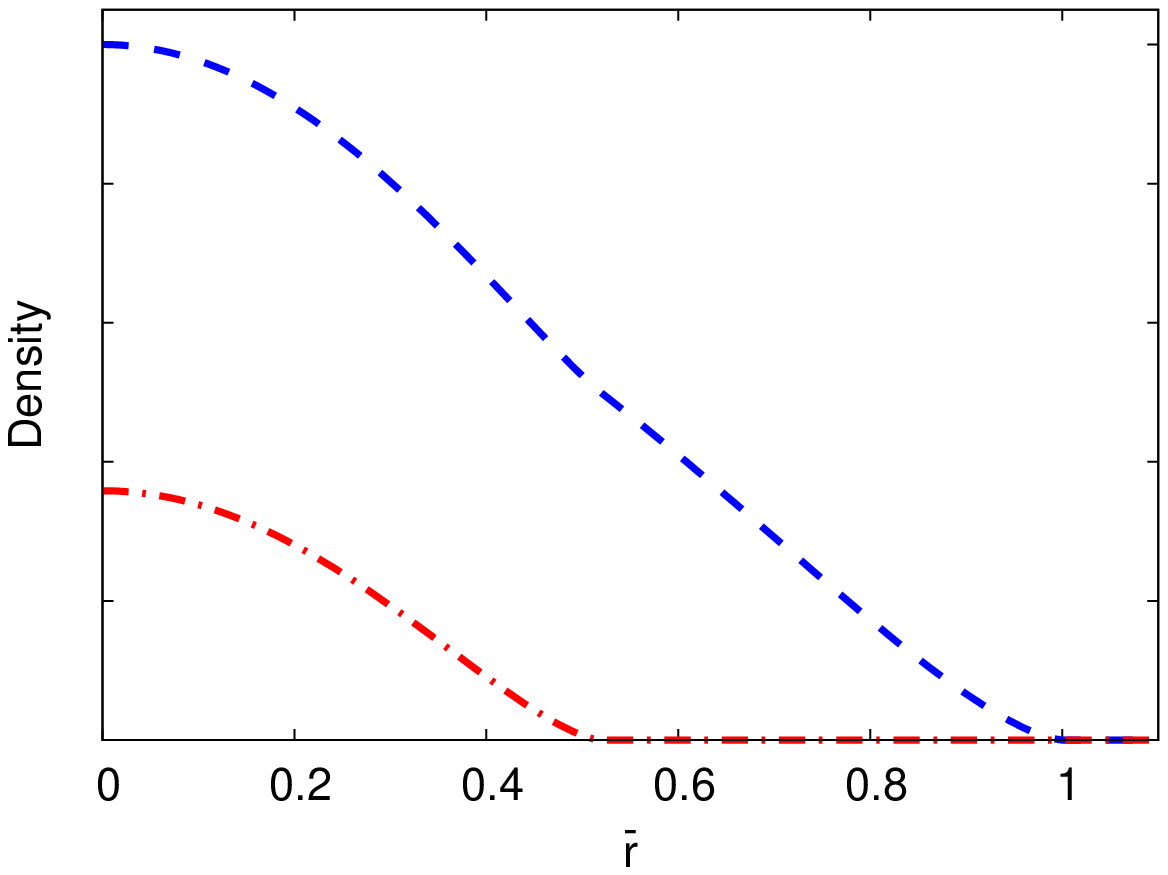}
\includegraphics[width=0.95\columnwidth]{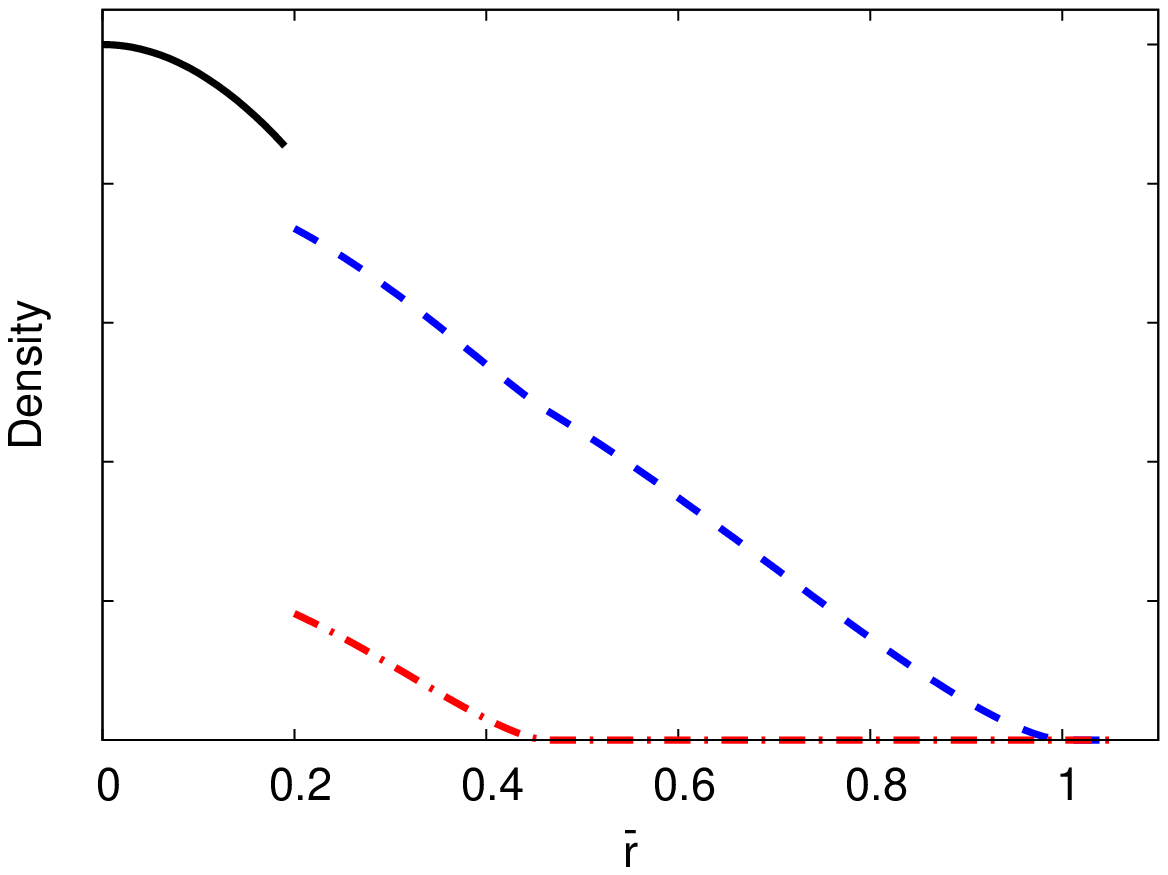}
\includegraphics[width=0.95\columnwidth]{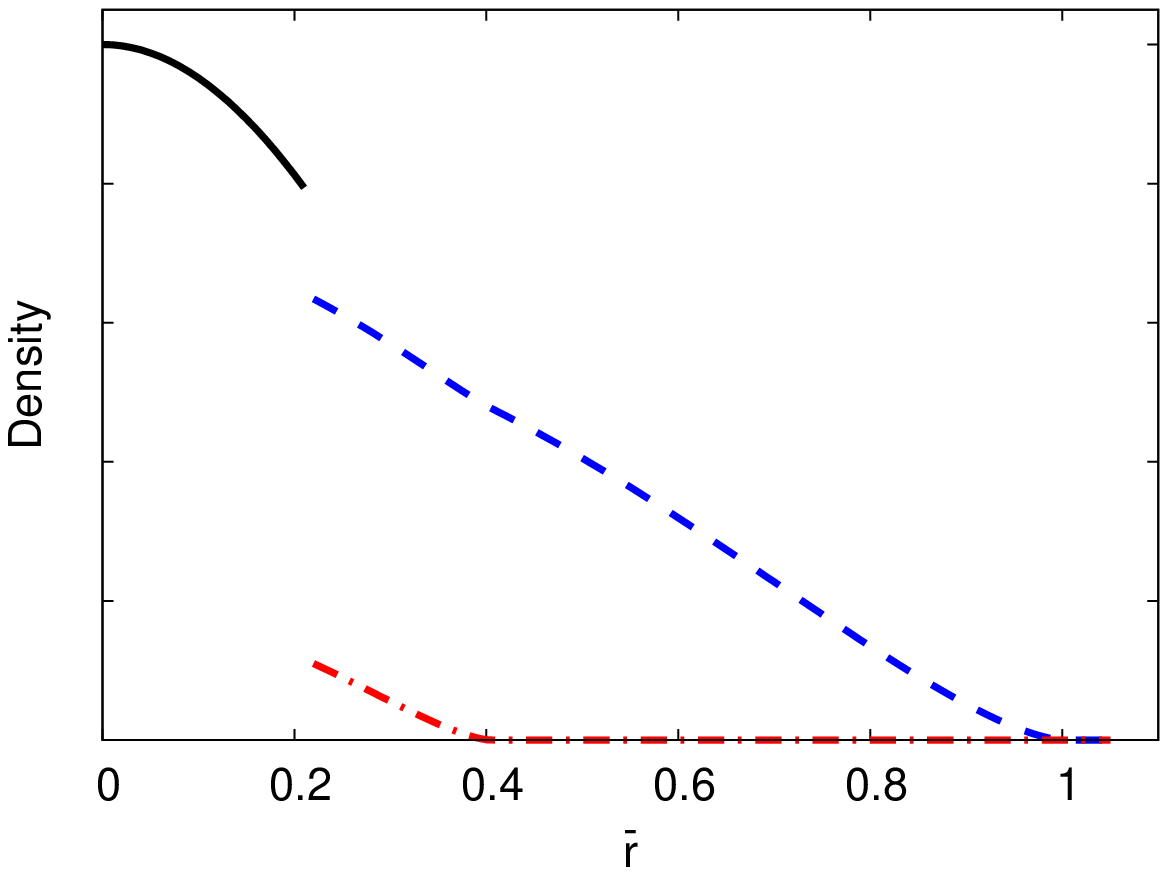}
\caption{\label{fig:6}(color online) Density profiles normalized by the density~$n_{\text{S}}$ at the center 
of the trap as a function of~$\bar{r}=r/R_{\uparrow}$ for fixed~$P=P_{\text{c},<}(\bar{m}=0.1)\approx 0.89$
and three different values of the mass-imbalance parameter~$\bar{m}=0.1, 0.2, 0.3$ (from top to bottom). 
The trap-asymmetry parameter has been set to~$\alpha=1$.
The (black) solid line depicts the density~$n_{\text{S}}$
in the superfluid regime, whereas the (blue) dashed and the (red) dashed-dotted 
lines depict~$n_{\uparrow}$ and~$n_{\downarrow}$, respectively.
}
\end{figure}

(a) The functional~\eqref{eq:edf} can be used to study the system for~$P\gtrsim P_{\text{c},<}(\bar{m})$ 
independent of our choice for~$\bar{m}$
but it does not allow us to describe reliably the system for~\mbox{$\bar{m}>\bar{m}_{\text{c},<}$} and~$P<P_{\text{c},<}(\bar{m})$, 
not even on a qualitative level. Therefore it should not be used
in this regime, at least it should not be considered for~$P\ll P_{\text{c},<}(\bar{m})$.

(b) The functional~\eqref{eq:edf} can be used to study reliably the system for~$\bar{m}>0$ but it is insufficient 
to describe the system for~$\bar{m}<0$. From our symmetry considerations, it would then follow that~$\bar{m}_{\text{c},<}\to -1$ 
as we would expect to have~$\bar{m}_{\text{c},>}\to 1$ in this case, see our discussion above. This implies that the underlying functional 
fails to predict the line of critical polarizations~$P_{\text{c},<}(\bar{m})$ for~$\bar{m}<0$. Note that the predictions from the 
functional~\eqref{eq:edf} for
the critical polarization and the density profiles for~$\bar{m}=0$ are in good agreement with MIT experiments, see 
also~Ref.~\cite{RLS}.

(c) The good agreement with MIT experiments for~$\bar{m}=0$ suggests that the functional~\eqref{eq:edf} should only
be used for studies with polarization~$P>{\mathcal P}(\bar{m})$. Here,~${\mathcal P}$ defines a lower bound for the applicability of the
functional~\eqref{eq:edf} which we expect to depend on~$\bar{m}$. From the comparison of the density profiles with those measured
in experiments, it moreover follows that, at least,~${\mathcal P}(0) \gtrsim 0.4$, see~Ref.~\cite{RLS}.

Our analysis appears to favor conclusion~(c). In fact, we
 find that the normalized equation of state~$\varepsilon(x)=(5/3)(\varepsilon_\text{N}(x)/\varepsilon_{\text{F} \uparrow})$
of the normal phase becomes negative for a finite range of values of~$x$ for~$\bar{m}\lesssim -0.5$ (see Fig.~\ref{fig:5}). As~$\varepsilon(x)$ enters 
our construction of the energy density functional~\eqref{eq:edf}, it follows that the latter 
can no longer be used to study trapped systems with~$\bar{m}\lesssim -0.5$.
Note that it may very well be that the functional~\eqref{eq:edf} becomes already unreliable for larger values of~$\bar{m}$ at~$P\gtrsim 0$. Therefore it is
reasonable to expect that our DFT does not allow for a quantitative prediction of the location of the ``critical point"~$\bar{m}_{\text{c},<}$. 
The reason for the breakdown
of our DFT approach for~$\bar{m}\lesssim -0.5$ can be manifold. As already discussed in Sect.~\ref{sec:formalismtr}, strictly speaking, 
our ansatz for~$\varepsilon(x)$ is only valid for~$x\ll 1$. For~$\bar{m}=0$, however, 
it has been found that our ansatz for~$\varepsilon(x)$ 
describes results from MC simulations very well, even for large values of $x$ (see Ref.~\cite{Lobo:2006}). 
For~$\bar{m}\neq 0$, this may no longer be the case as,
for example, higher-order terms may become relevant. Moreover, three-body effects may become important in this regime. The relevance of 
the latter effects has also been pointed out by analytic studies of 
few-body systems (see, e.g., Refs.~\cite{Braaten:2004rn,Nishida:2007mr,Niemann:2012gd}), as well as by
Quantum Monte-Carlo studies~\cite{Gandolfi:2010}. 
In any case, an improvement of the equation of state of the normal phase in this direction is beyond the scope of
the present work. Still, our analysis of the phase diagram of the uniform system as well as of the symmetries of the theory suggests
the existence of two ``critical points"~$\bar{m}_{\text{c},<}$ and~$\bar{m}_{\text{c},>}$ at~$P=0$ 
with~$\bar{m}_{\text{c},>}=-\bar{m}_{\text{c},<}$ and~$\bar{m}_{\text{c},>}<1$. However, a 
computation of the precise values of these points is not possible with the energy density functional underlying our present work.

\subsubsection{Density profiles}

Finally, we would like to briefly discuss the density profiles as obtained from our DFT study. To this end, we restrict ourselves to the case of large values
of the polarization~$P$ where, following our discussion above, we still expect our present ansatz for the energy density functional to yield 
reliable results. In Fig.~\ref{fig:6}, we show the density
profiles for fixed~$P=P_{\text{c},<}(\bar{m}=0.1)\approx 0.89$ and three different values of the mass-imbalance parameter~$\bar{m}=0.1, 0.2, 0.3$. We find that
the ratio~$n_{\uparrow}/n_{\text{S}}$ evaluated at~$r=R_{\text{S}}$ decreases 
with increasing~$\bar{m}$.\footnote{Note that~$n_{\uparrow}/n_{\text{S}}\approx 1$
for~$\bar{m}=0$ and~$P\lesssim P_{\text{c},<}(0)\approx 0.68$ (see also Ref.~\cite{RLS}).} 
The decrease of~$n_{\uparrow}/n_{\text{S}}$ at~$r=R_{\text{S}}$
with~$\bar{m}$, together with our predictions for the density profiles 
themselves, can be viewed as a testable prediction for future experimental studies of
mass- and spin-imbalanced unitary Fermi gases. In the same way, we expect that our result for 
the line of critical polarizations~$P_{\text{c},<}(\bar{m})$ can provide reliable guidance for 
experiments with~$\bar{m}\gtrsim 0$.

\section{Summary}\label{sec:sum}
In this work we have studied the phase diagram of trapped mass- and spin-imbalanced unitary Fermi gases.
To this end, we constructed an energy density functional along the lines of Ref.~\cite{RLS}. This allowed us to
compute the critical polarization as a function of the mass-imbalance parameter~$\bar{m}$, at least for strongly
spin-imbalanced systems. 
On the other hand, our symmetry considerations together with our results for the phase diagram of the uniform
system strongly suggest that the energy density functional underlying our studies is insufficient to reliably study
trapped mass-imbalanced Fermi gases for small spin-polarizations~$P$. Nevertheless, our study allowed us
to understand the structure of the phase diagram on a qualitative level. 

Our analysis suggests that, in addition to the line of critical polarizations predicted by our DFT approach, a second line of critical polarizations exists.
Moreover, our analysis suggests that
the emergence of a superfluid region at the center of the trap is reasonably well described by our energy density
functional, provided that we study the case of (highly) spin-polarized systems for~$\bar{m}\gtrsim 0$. 
In this regime, we also expect that our predictions for the density profiles are meaningful.
For~$\bar{m}\lesssim 0$, we find that the critical polarization associated with the emergence of a superfluid
region in the center of the trap tends to zero. Whereas this behavior of the critical polarization is reasonable and 
can be understood on more general grounds, we believe that our present ansatz for the energy density functional is not capable of
predicting accurately the value of~$\bar{m}<0$ at which the critical polarization vanishes.

An improvement of our present DFT study requires a detailed analysis of the equation of state of the normal phase,
see Eq.~\eqref{eq:eosn} and possibly also of the superfluid phase. To this end, it may very well be required to 
study the equation of state of a
homogeneous, mass- and spin-imbalanced gas with various different non-perturbative approaches, such as lattice MC
calculations~(see, e.g., Ref.~\cite{Drut:2008} for a review), Renormalization Group 
approaches (see, e.g., Refs.~\cite{Diehl:2009ma,Bartosch:2009zr,Scherer:2010sv,Braun:2011pp,Boettcher:2012cm}), 
and two-particle irreducible approaches (see, e.g., Ref.~\cite{Haussmann:2007zz}). In this respect, the role of three-body effects
should possibly also be taken into account. Such advanced studies of the  
equation of state may also help to analyze in which region of the phase diagram, and to what extent, 
we may see ``signals" of inhomogeneous phases in experimental data.
 

\acknowledgments
The authors thank F. Chevy and R. Grimm for helpful discussions.
J.B. and D.R. acknowledge support by the DFG under Grant BR 4005/2-1. Moreover, the authors acknowledge support
by HIC for FAIR within the LOEWE program of the State of Hesse. J.E.D. acknowledges funding from the U.S. National Science Foundation,
under grant No. PHY1306520.

\appendix
\section{Mean-field study of a uniform spin- and mass-imbalanced Fermi gas}\label{app:mf}
In this appendix we briefly discuss the mean-field study underlying our discussion of the phase diagram of the
uniform system in Sect.~\ref{sec:formalismuf}. In order to compute the latter phase diagram, we have derived 
the order-parameter potential~$U$ for U($1$) symmetry breaking in the mean-field approximation from the path intergral
representation of the partition function~$\mathcal Z$ (see Refs.~\cite{Braun:2012ww,Roscher:2013aqa} 
and Refs.~\cite{ChevyMora,StoofGubbels} for more general reviews).
In the unitary limit, we obtain
\be
&&\beta U(\bar{\varphi}) = -2\beta {\mu} |\bar{\varphi}|^2 - \int\frac{d^3{q}}{(2\pi)^3}\ln\left[ \cosh\left(\beta\bar{m} q^2 + \beta h\right)\right. \nn \\
&& \qquad\qquad\qquad \left. + \cosh\left(\beta\sqrt{\left(q^2 - {\mu}\right)^2 + g_\varphi^2|\bar{\varphi}|^2}\right)\right]\,,\label{eq:mfu}
\ee
where the background field (mean field)~$\bar{\varphi}=\varphi-\delta\varphi$ is defined to be the difference of the auxiliary field~$\varphi\sim g_{\varphi}\psi_{\uparrow}\psi_{\downarrow}$
and the fluctuation field~$\varphi$.
In this work, we do not take into account that the ground-state 
configuration may break translation invariance, even in the uniform system (see, e.g., Ref.~\cite{Roscher:2013cma}).
Note that we have dropped standard $\varphi$-independent terms in Eq.~\eqref{eq:mfu} which are required to regularize the potential.

The order-parameter potential~$U$ and the grand canonical potential~$\Omega$ are related,~$\Omega=VU(\varphi^{}_0)$, where~$V$ 
is the volume of the system and~$\varphi^{}_0$ denotes the value of $\varphi$ minimizing the potential. Moreover, $g_{\varphi}^2 |\varphi^{}_0|^2$
can be identified with the fermion gap~$\Delta$ which serves as an order parameter for spontaneous U($1$) symmetry breaking 
associated with a superfluid ground state. 

Our results for dimensionless (universal) quantities extracted from the potential~$U$, such as the {\it Bertsch} 
parameter, are independent of our choice for~$\mu$ and 
the coupling~$g_{\varphi}$ of the fermions to the (auxiliary) field~$\varphi$. 
Note that the four-fermion coupling is directly related to the coupling~$g_{\varphi}$. In fact, the latter is chosen 
to reproduce the four-fermion term in the action associated with the Hamilton operator defined in Sect.~\ref{sec:formalismuf}.

\bibliographystyle{h-physrev3}
\bibliography{bib_source}

\end{document}